\documentclass[onecolumn,floatfix,superscriptaddress,a4paper,
               showpacs,showkeys,nofootinbib,preprint]{revtex4}
\usepackage{epsfig}%
\usepackage{latexsym}
\usepackage{xspace}
\usepackage{hyperref}
\usepackage[latin2]{inputenc}
\usepackage{indentfirst}
\usepackage{enumerate}

\usepackage{amsmath}
\usepackage{amssymb}
\usepackage[english]{babel}
\usepackage{url}

\newcommand{\eq}[1]{\begin{align} #1 \end{align}}

\begin{document}
\title{Hadron multiplicities and chemical freeze-out conditions\\ in proton-proton and nucleus-nucleus collisions
%:\\ comparison of the data with statistical and transport models
}

\author{V. Vovchenko}
\affiliation{
Frankfurt Institute for Advanced Studies, Johann Wolfgang Goethe University, D-60438 Frankfurt, Germany}
\affiliation{
Taras Shevchenko National University of Kiev, 03022 Kiev, Ukraine}
\author{V. V. Begun}
\affiliation{Institute of Physics, Jan Kochanowski University, PL-25406 Kielce, Poland}
\author{M. I. Gorenstein}
\affiliation{Bogolyubov Institute for Theoretical Physics, 03680 Kiev, Ukraine}
\affiliation{
Frankfurt Institute for Advanced Studies, Johann Wolfgang Goethe University, D-60438 Frankfurt, Germany}

\begin{abstract}
New results of the NA61/SHINE Collaboration at the CERN SPS
on mean hadron multiplicities in proton-proton (p+p) interactions are analyzed
within the transport models and the hadron resonance gas (HRG)
statistical model.
The chemical freeze-out parameters
in p+p interactions and central Pb+Pb (or Au+Au)
collisions are found and compared
with each other in the range of the center of mass energy
of the nucleon pair $\sqrt{s_{NN}}=3.2-17.3$~GeV.
The canonical ensemble formulation of the HRG model is used to
describe mean hadron multiplicities  in p+p interactions and the grand
canonical ensemble  in central Pb+Pb and Au+Au
collisions. The chemical freeze-out temperatures
in p+p interactions are found to be larger than the corresponding
temperatures in central nucleus-nucleus collisions.
\end{abstract}



\maketitle

\section{Introduction}
%A search for the quark gluon plasma (QGP)
Studies of properties of the strongly-interacting matter at
extreme energies and densities is one of the main goals for
high-energy nucleus-nucleus (A+A) collision experiments.
The data of the NA49 Collaboration on hadron production in central
Pb+Pb collisions~\cite{NA49-1,NA49-3,NA49-2} at beam energies
$E_{\rm lab}=$20$A$, 30$A$, 40$A$, 80$A$, and 158$A$~GeV (which
corresponds to $\sqrt{s_{NN}}=6.3,~7.6,~8.8,~12.3,~17.3$~GeV for
the center of mass energy of the nucleon pair) show rapid changes
of several hadron production properties. Particularly, the sharp
maximum of the $K^+/\pi^+$ ratio (the {\it horn}) at 30$A$~GeV
predicted in Ref.~\cite{horn} was found with approximate constant
value of this ratio at collision energies higher than 80$A$~GeV.
These data were obtained at the Super Proton Synchrotron (SPS) of
the European Organization for Nuclear Research (CERN), and they
have been confirmed by the Beam Energy Scan (BES)
program~\cite{RHIC} at the Relativistic Heavy Ion Collider (RHIC)
at the Brookhaven National Laboratory (BNL).

These results have different interpretations. For instance, the
results are consistent with the onset of deconfinement in central
Pb+Pb collisions at about 30$A$~GeV \cite{GaGo}, assuming that
hadron production is mainly determined by the properties of the
early stage of the collision. On the other hand, the strangeness
horn is also %adequately
qualitatively described within the thermal
model~\cite{BraunMunzinger:2001as,Cleymans:2016qnc}, especially
when some modifications are considered,
%missing only the description of the tip of the horn,
see e.g.
Refs.~\cite{Andronic:2008gu,Oliinychenko:2012hj,Naskret:2015pna}.
Therefore, these data do not allow to make firm conclusions with
regard to the onset of deconfinement. The successor of the NA49,
the NA61/SHINE Collaboration, is performing the scan of the beam
energy and system size at the
SPS~\cite{Ga:2009,NA61facility,NA61-p+p-mult}. Additionally, the
BES program at RHIC~\cite{RHIC} studies Au+Au collisions in the
energy range of $\sqrt{s_{\rm NN}} = 7.7-200$~GeV.
%The aim of
%these studies is to search for the critical point of strongly
%interacting matter and investigate properties of the onset of
%deconfinement.
An important aspect of these studies is a comparison of A+A and
p+p collisions.
%: it is usually  assumed that the QGP can be formed
%in A+A, but not in p+p reactions.
Information about the physical
properties of the system created in Pb+Pb and p+p collisions
is also very useful in a sense that these reactions represents two limiting
cases of the system-sizes.
%  dependence.
%{\bf In
%this respect the new results of the NA61/SHINE Collaboration on
%mean hadron multiplicities in p+p inelastic interactions look
%rather intriguing, because instead of the horn in A+A they show
%the step-like structure at the same energies in p+p, see
%also~\cite{Poberezhnyuk:2015wea}.
%thus,
%the physical properties for these two systems at given collision energy

%provides a baseline for collisions of lighter nuclei.
%The
%scanning over the system size should answer the question whether
%the deconfinment signals will be also seen in collisions of light
%nuclei and  p+p interactions.

In the present paper we analyze the new p+p data of the
NA61/SHINE at $p_{\rm lab}=$~20, 31, 40, 80, and
158~GeV/c~\cite{NA61-p+p-mult} (which corresponds to
$\sqrt{s_{NN}}=6.3,~7.7,~8.8,~12.3,~17.3$~GeV). We also repeat the
analysis of the NA49 data in central Pb+Pb collisions. This is
done because of an extension  of the NA49 data on central Pb+Pb
collisions~\cite{NA49-1,NA49-3,NA49-2,NA49-latest,NA49-4} in
comparison to the results used in previous studies, e.g., in
Ref.~\cite{Bec}.

We analyze also the recent data from the HADES Collaboration for
both p+p collisions at $E_{\rm kin}=3.5~$GeV~\cite{HADES-p+p}, and
Au+Au collisions at $E_{\rm kin}=1.23A~$GeV~\cite{HADES-A+A}. The
corresponding center of mass energies are $\sqrt{s_{NN}}=3.2~$GeV
for p+p and $2.4~$GeV for A+A.
For completeness of the analysis we redo the early fits of the
Au+Au collisions for $E_{\rm kin}=0.8A,~1.0A~$GeV at GSI
Schwerionensynchroton (SIS), and for the $E_{\rm lab}=11.6A~$GeV
at BNL Alternating Gradient Synchrotron
(AGS)~\cite{Cleymans:1998yb,Averbeck:2000sn,AGS1,AGS2,Becattini:2000jw}. The
corresponding center of mass energies are
$\sqrt{s_{NN}}=2.2,~2.3,$ and 4.9~GeV.

Therefore, the analyzed energy range is
$\sqrt{s_{NN}}=3.2-17.3$~GeV for p+p interactions and
$\sqrt{s_{NN}}=2.2-17.3$~GeV for central Pb+Pb or Au+Au
collisions.
Such an analysis extends previous studies
regarding the systematic comparison of the
hadron production properties in A+A and p+p collisions
to energies below $\sqrt{s_{NN}}=17.3$~GeV.
It should be noted that there are two
facilities under construction which will operate  in the considered energy
region:
%in order to make studies in this energy range:
the Facility for Antiproton and Ion Research
(FAIR)~\cite{CBMPhysicsBook,cbm}, % at GSI
and the Nuclotron-based Ion Collider fAcility (NICA)~\cite{NICA}.
%at Joint Institute for Nuclear Research (JINR) in Dubna.

The data on hadron multiplicities are compared
with predictions of two popular transport models -- Ultra-relativistic Quantum
Molecular Dynamics (UrQMD)~\cite{UrQMD,UrQMD:2008,UrQMD:2014} and
Hadron String Dynamics (HSD)~\cite{HSD1,HSD2,HSD3}.
The properties of p+p interactions are the input
to these models. Therefore, we test whether
%how well
this input obtained from the parametrization of previous p+p
results allows to reproduce the new NA61/SHINE data.

We perform the fits of the mean hadron multiplicities within
statistical Hadron Resonance Gas (HRG) model in the Grand
Canonical Ensemble (GCE) for central Pb+Pb and Au+Au collisions,
except for the lowest energies at SIS where the
strangeness-canonical ensemble (SCE) was used, and in the
Canonical Ensemble (CE) for p+p inelastic interactions.
These fits provide us with the HRG model parameters as the
functions of collision energy. Particularly, we present a
comparison of the chemical freeze-out temperatures in p+p and
central collisions of heavy ions in the energy region
$\sqrt{s_{NN}}=3.2-17.3$~GeV. In the literature, such a comparison
was previously limited to higher collision energies
$\sqrt{s_{NN}}\geq 19.4$~GeV for p+p~\cite{Becattini:1997rv}. Our
results enable one to estimate the range of the chemical
freeze-out parameters that can be reached in collisions of
different size nuclei during the current energy and system
size scanning by NA61/SHINE at SPS, and in the future experiments
at NICA and FAIR.

The paper is organized as follows.  In Sec.~\ref{transp} the
results of the calculations of mean hadron multiplicities in
inelastic p+p reactions within the UrQMD and HSD models are
presented. The HRG is considered in Sec.~\ref{hrg}. The results
for p+p inelastic interactions and central A+A collisions are
presented. The summary in Sec.~\ref{sum} closes the paper.

\section{Transport Models for Proton-Proton Collisions }\label{transp}

In this section the UrQMD and HSD transport models results
regarding inelastic p+p interactions are compared
with the NA61/SHINE data  \cite{NA61-p+p-mult}
\begin{figure}
%\centering
\includegraphics[width=0.49\textwidth]{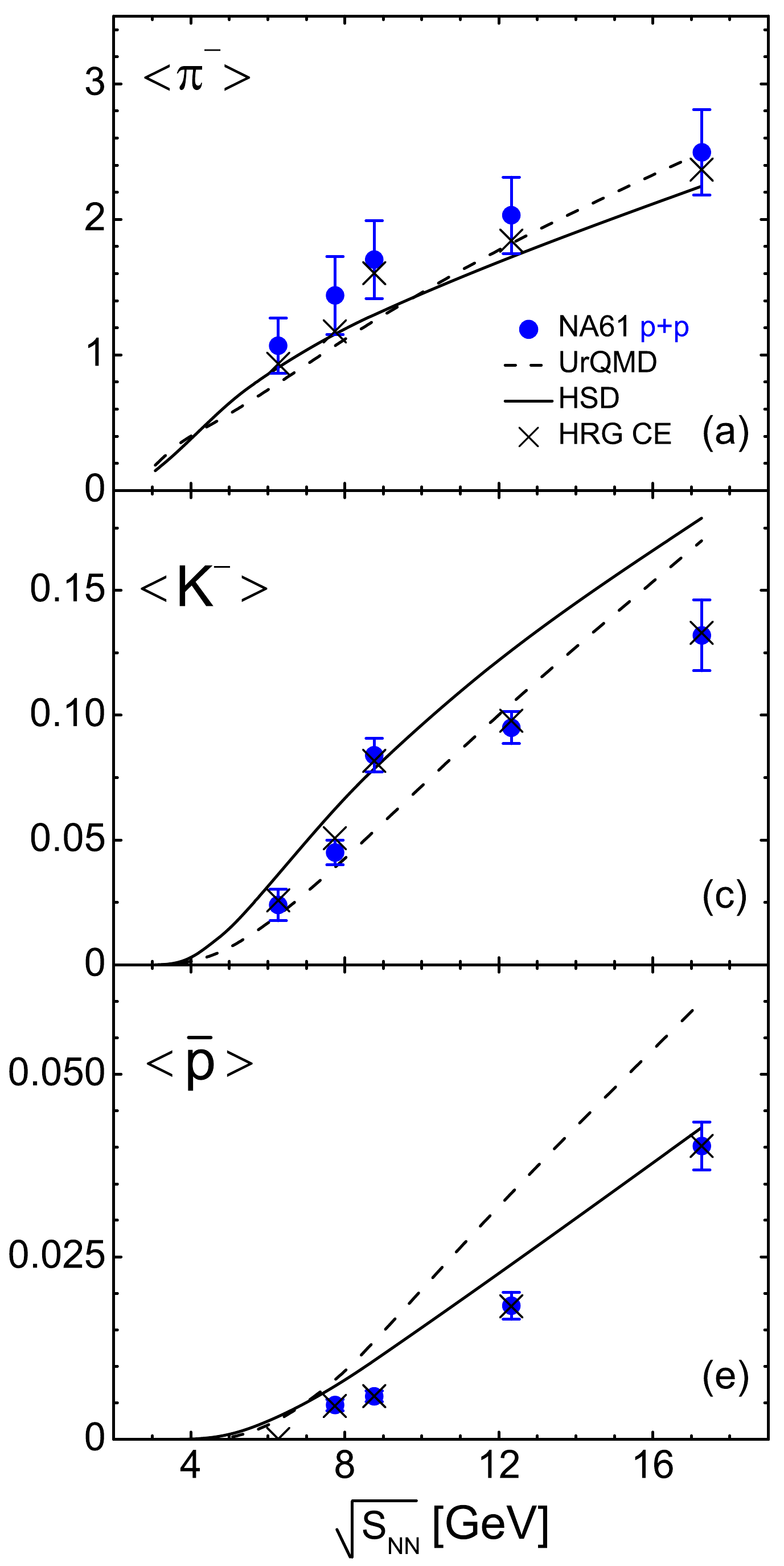}~~
\includegraphics[width=0.49\textwidth]{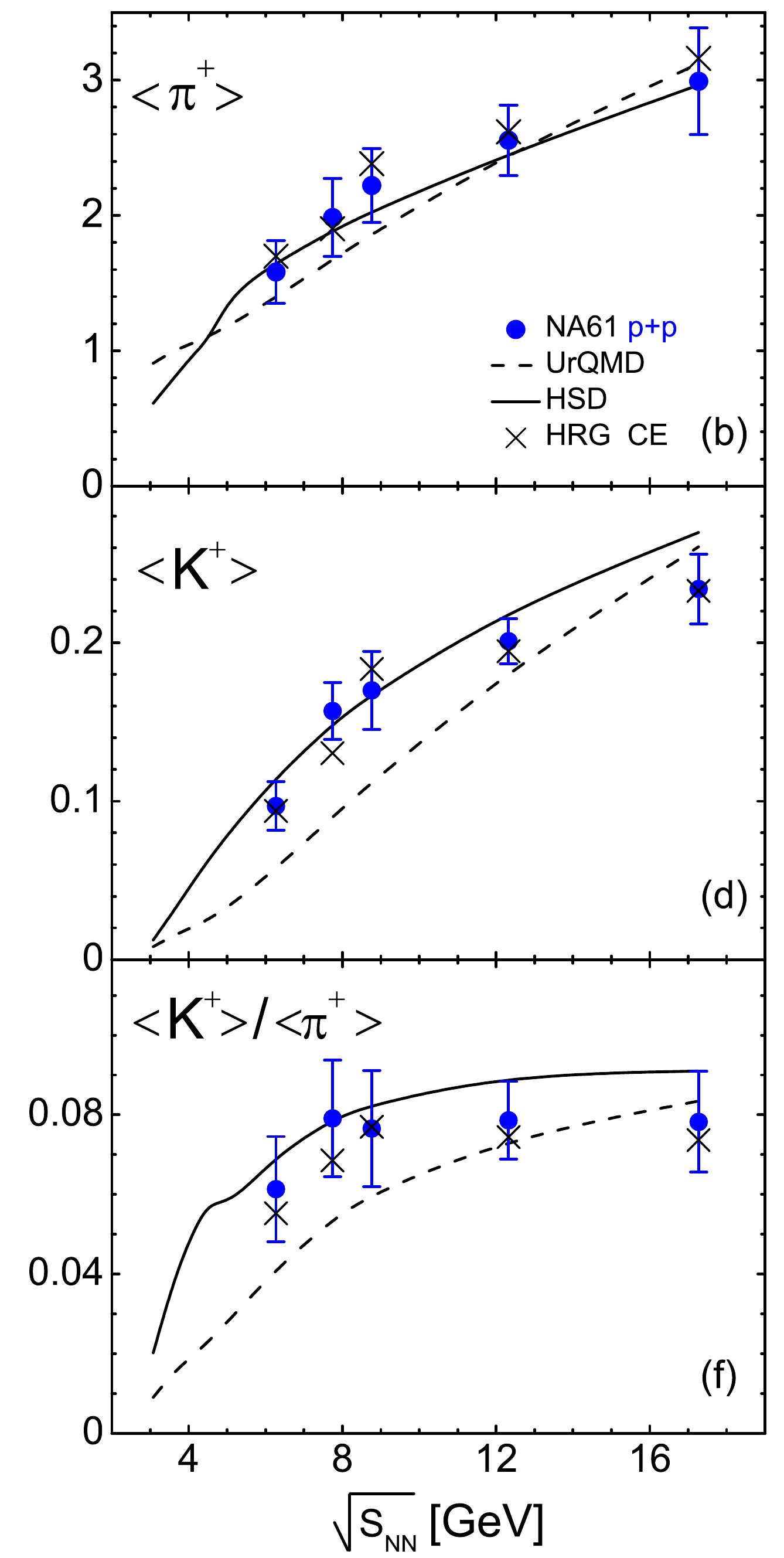}
\caption{Mean hadron multiplicities in inelastic p+p reactions as
functions of the center of mass collision energy $\sqrt{s_{NN}}$.
The full circles are the data \cite{NA61-p+p-mult} of NA61/SHINE
Collaboration. The solid and dashed lines correspond to the
results of the HSD 2.5 and UrQMD 3.4 model simulations,
respectively. The crosses show the fit in the statistical
hadron-resonance gas model in the canonical ensemble (see
Sec.~\ref{hrg}). }\label{fig-pi}
\end{figure}
%
%The data of the NA61~\cite{NA61-p+p-mult} contain the 4$\pi$
%multiplicities
on mean hadron multiplicities  $\langle \pi^+\rangle$, $\langle \pi^-\rangle$,
$\langle K^+\rangle $, $\langle K^-\rangle $,
and $\langle \overline{p}\rangle $ at $p_{\rm lab}=$~20, 31, 40, 80, 158~GeV/c
which corresponds to $\sqrt{s_{NN}}=6.3,~7.7,~8.8,~12.3,~17.3$~GeV
(the $\overline{p}$ multiplicity was not reported for the
lowest collision energy).
A comparison of the NA61/SHINE results with UrQMD and HSD predictions
is presented in Figs.~\ref{fig-pi} (a)-(e).
In addition, the $\langle K^+\rangle/\langle\pi^+\rangle$ ratio is
shown in Fig.~\ref{fig-pi} (f).
From Fig.~\ref{fig-pi} (a) one concludes that both transport
models -- UrQMD and HSD -- underestimate yields of $\langle
\pi^-\rangle $ at all SPS energies, except the highest one,
$\sqrt{s_{NN}}=17.3$~GeV. As seen from Figs.~\ref{fig-pi} (d) and
(e) the UrQMD also underestimates $\langle K^+\rangle $ and
overestimates $\langle \overline{p}\rangle$. The both models
have problems describing the yields of $K^-$ shown in
Fig.~\ref{fig-pi} (c). The $\pi^-$ momentum spectra in p+p
reactions within the UrQMD have been analyzed and compared to
properly normalized $\pi^-$ spectra in central Pb+Pb collisions in
Ref.~\cite{VAG} (see also Ref.~\cite{mT}, where the mean values of
hadron transverse mass have been calculated within the UrQMD and
HSD models).
It should be noted that the p+p results are used as the input for the
transport model description of A+A collisions.
Therefore, it is evident that improvements of
parametrization of p+p results in both the UrQMD and HSD models
are really needed.

\section{Hadron Resonance Gas Model}\label{hrg}
\subsection{The model formulation}\label{subsec:themodel}
Statistical models appear to be rather successful in
calculations of mean hadron multiplicities in high energy collisions.
This approach assumes a thermodynamical equilibrium of stable hadrons
and resonances at the chemical freeze-out state described
%by a few unknown parameters which should be found
by thermal parameters to be determined
by fitting data.
A general description of the HRG model can be found elsewhere,
e.g., in the introduction part of Ref.~\cite{THERMUS}.
%
%The HRG model can be formulated in different statistical ensembles.

In the GCE formulation of the HRG the conserved charges, such as
baryonic number $B$, electric charge $Q$, and net strangeness $S$,
are conserved on average, but can differ from one microscopic
state to another. In the CE formulation these charges are fixed to
their exact conserved values in each microscopic state. The
distinct difference appears between calculations of hadron
multiplicities in different statistical ensembles, if the number
of particles with corresponding conserved charge is of the order
of unity or smaller
\cite{CE,CE1,CE2,Becattini:1997rv,CE3,Begun:2005qd}. In the
considered range of collision energies the CE is relevant for p+p
collisions, while GCE can be used for central Pb+Pb and Au+Au
collisions, except for the lowest energies at SIS. The exact
conservation of net strangeness needs to be enforced there, i.e.,
the calculations for these low-energy A+A collisions are done
within the SCE~\cite{BraunMunzinger:2001as,Cleymans:2016qnc}.

In the GCE the fitting parameters are the temperature $T$,
baryonic chemical potential $\mu_B$\footnote{The chemical
potentials $\mu_S$ and $\mu_Q$ correspond to the conservation of
strangeness and electric charge, respectively.
%in strong interactions.
They are found from the conditions of zero net strangeness and
fixed proton to neutron ratio in the colliding nuclei.}, the
system volume $V$, and the strangeness under-saturation parameter
$\gamma_S$ which is discussed in Ref.~\cite{Rafelski:2015cxa}. For
the convenient comparison between A+A and p+p we use the radius
$R$ calculated from $V\equiv 4\pi R^3/3$ instead of volume.
The CE treatment of p+p collisions assumes the fixed values of the
conserved charges -- baryonic number $B=2$, electric charge $Q=2$,
and net strangeness $S=0$. Therefore,
%one has no $\mu_B$
in the CE the fitting parameters are $T$, $R$, and $\gamma_S$. In
the SCE for SIS we set the strangeness conservation radius equal
to the radius of the system. Therefore, the fit parameters in the
SCE are the same as in the CE, but only strangeness is conserved
exactly. Note that at low collision energies a role of the exact
energy conservation becomes quite important. One should then
follow the micro canonical ensemble formulation which has not been
used in the present paper.

The HRG
model fits are  done by minimizing the value
 \eq{\label{xi}
 \frac{\chi^2}{N_{\rm dof}}
 ~=~\frac{1}{N_{\rm dof}}\sum_{i=1}^N\frac{\left(N_i^{\rm exp}~-~N_i^{\rm HRG}\right)^2}{\sigma_i^2}~,
}
where $N_i^{\rm exp}$ and $N_i^{\rm HRG}$ are the experimental and
calculated in the HRG hadron multiplicities, respectively; $N_{\rm
dof}$ is the number of degrees of freedom, that is the number of
the data points minus the number of fitting parameters; and
$\sigma_i^2=(\sigma_i^{syst})^2+(\sigma_i^{stat})^2$ is the sum of
the squares of the statistical and systematic experimental errors.

In our calculations we include stable hadrons and resonances that
are listed by the Particle Data Group~\cite{pdg}, and take into
account both the quantum statistics, and the Breit-Wigner shape of
resonances with finite widths. The list of particles includes
mesons up to $f_2(2340)$, (anti-)baryons up to $N(2600)$, and
generally corresponds to the newest THERMUS~3.0~\cite{THERMUS}
compilation. We do not include hadrons with charm and bottom
degrees of freedom which have a negligible effect on the fit
results. {In contrast to the
Refs.~\cite{Andronic:2008gu,Oliinychenko:2012hj} we} also removed
the $\sigma$ meson ($f_0(500)$) and the $\kappa$ meson
($K_0^*(800)$) from the particle list because of the reasons
explained in
Refs.~\cite{GomezNicola:2012uc,Venugopalan:1992hy,Broniowski:2015oha,Pelaez:2015qba}.

The mean multiplicity $\langle N_i\rangle$ of $i$th
particle species  is calculated in the HRG model
as a sum of the primordial mean multiplicity
$\langle N^{\rm prim}_i\rangle$ and resonance decay
contributions as follows
\eq{\label{eq:Ntot}
\langle N_i\rangle~ =~
\langle N^{\rm prim}_i\rangle~ +~ \sum_R \langle n_i \rangle_R \, \langle N^{\rm prim}_R\rangle~,
}
where $\langle n_i \rangle_R$ is the average number of particles
of type $i$ resulting from decay of resonance $R$.
Note that Eq.~\eqref{eq:Ntot} is also valid for calculating yields
of unstable particles, such as the $\phi$ meson, $K^*(892)$
resonance, or $\Lambda(1520)$ resonance. This is important since
yields of these  unstable particles have been measured (see, e.g.
Ref.~\cite{HADES-p+p,NA49-4,NA49-latest,NA49-p+p}). Note, however,
that the present version of THERMUS does not take into account the
resonance decay contribution to mean multiplicities of particles
which are marked as unstable. As a result, yields of $\phi$,
$K^*(892)$, or $\Lambda(1520)$ can be underestimated by up to
25$\%$. The actual amount depends on the HRG parameters used, and
on the modeling of relevant decay branching ratios, which are
sometimes poorly constrained. On the other hand, if, e.g., one
marks the $\phi$ as a stable particle in THERMUS, then the decay
contribution to the $\phi$ multiplicity is calculated, but the
further decays of $\phi$ to kaons or pions are not taken into
account in the program, while they are accounted in the
experiment. To avoid this problem and to simultaneously fit
yields of stable and unstable hadrons in THERMUS one has to use
multiple particle sets.
%the standard THERMUS version does not permit to calculate
%correctly multiplicities of resonances, and to take into account
%their decays simultaneously.
%
%{\bf Perhaps, the easiest way to see it, is to mark some stable
%hadron, say $\pi^+$, as unstable, and compare results. One would
%expect that such modification does not lead to any changes, since
%pions do not decay into other hadrons. However, the total $\pi^+$ yield output by
%THERMUS is much smaller in this case, it consists of only thermal
%pions, and no longer includes any contribution from resonance
%decays.}
%
%To avoid this problem,
Alternatively, one can add
an extra loop for the summation of the
decay contributions to the yields of unstable particles in the
THERMUS code.

We have verified that in this case THERMUS yields
essentially the same results for total hadron yields of all
particles as our own implementation of the HRG. Thus, we use the
latter in all our subsequent analysis. We also enable the
calculation of asymmetric error bars for the obtained parameters,
which are obtained by explicitly analyzing the $\chi^2 =
\chi^2_{\rm min} + 1$ contours.

\subsection{HRG results for central A+A collisions and p+p inelastic reactions}

\begin{figure}[ht!]
\centering
\includegraphics[width=0.49\textwidth]{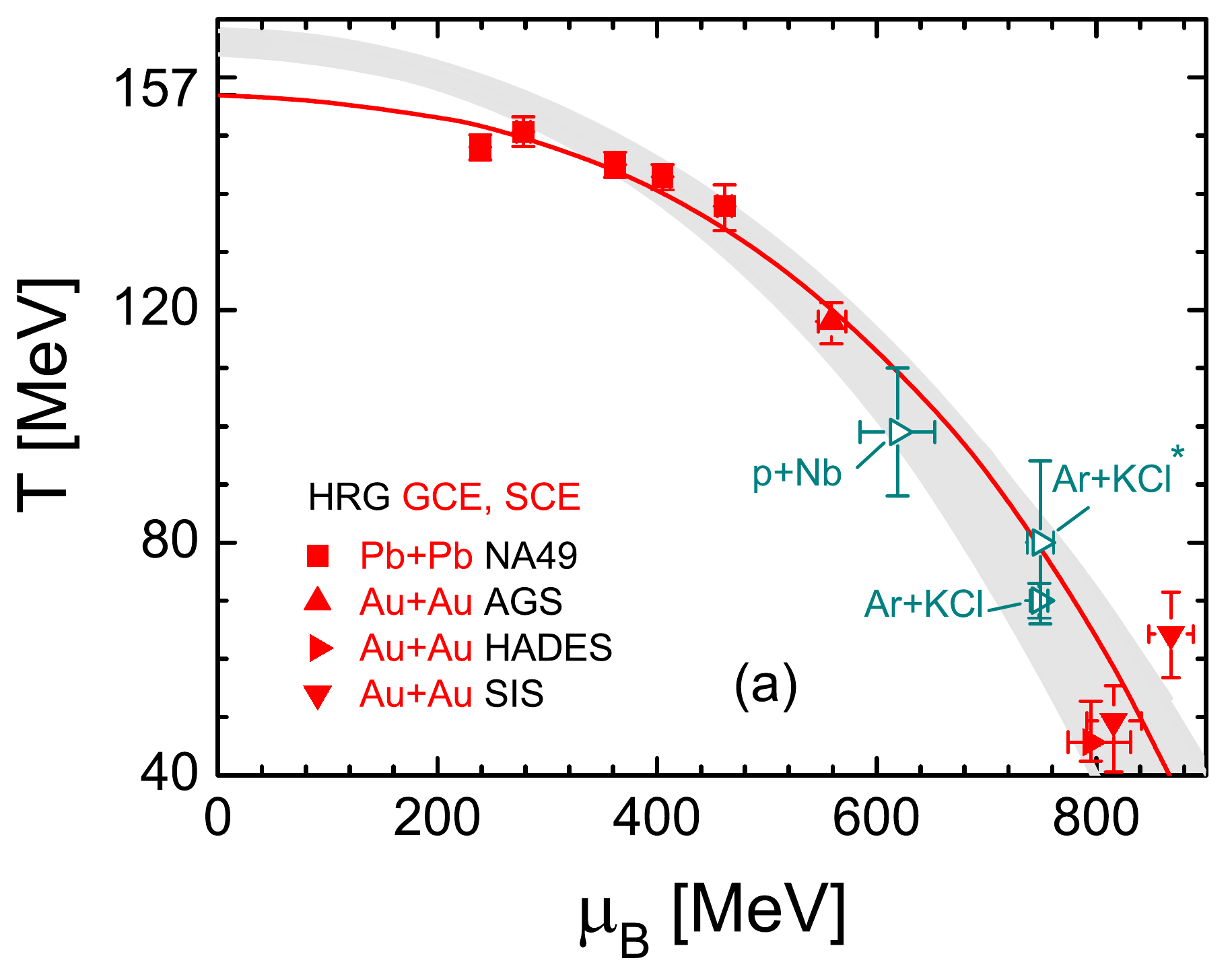}
\includegraphics[width=0.49\textwidth]{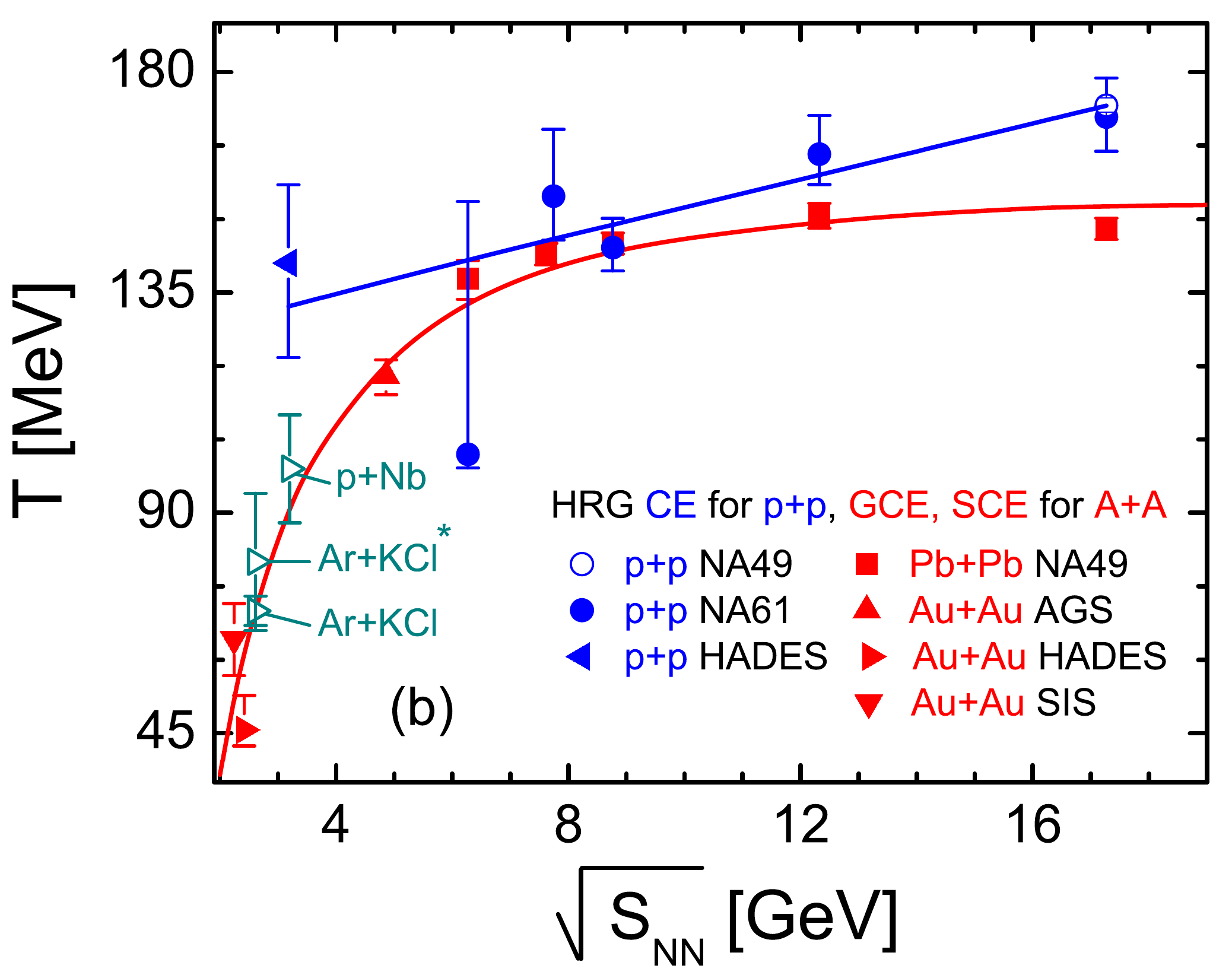}
\includegraphics[width=0.49\textwidth]{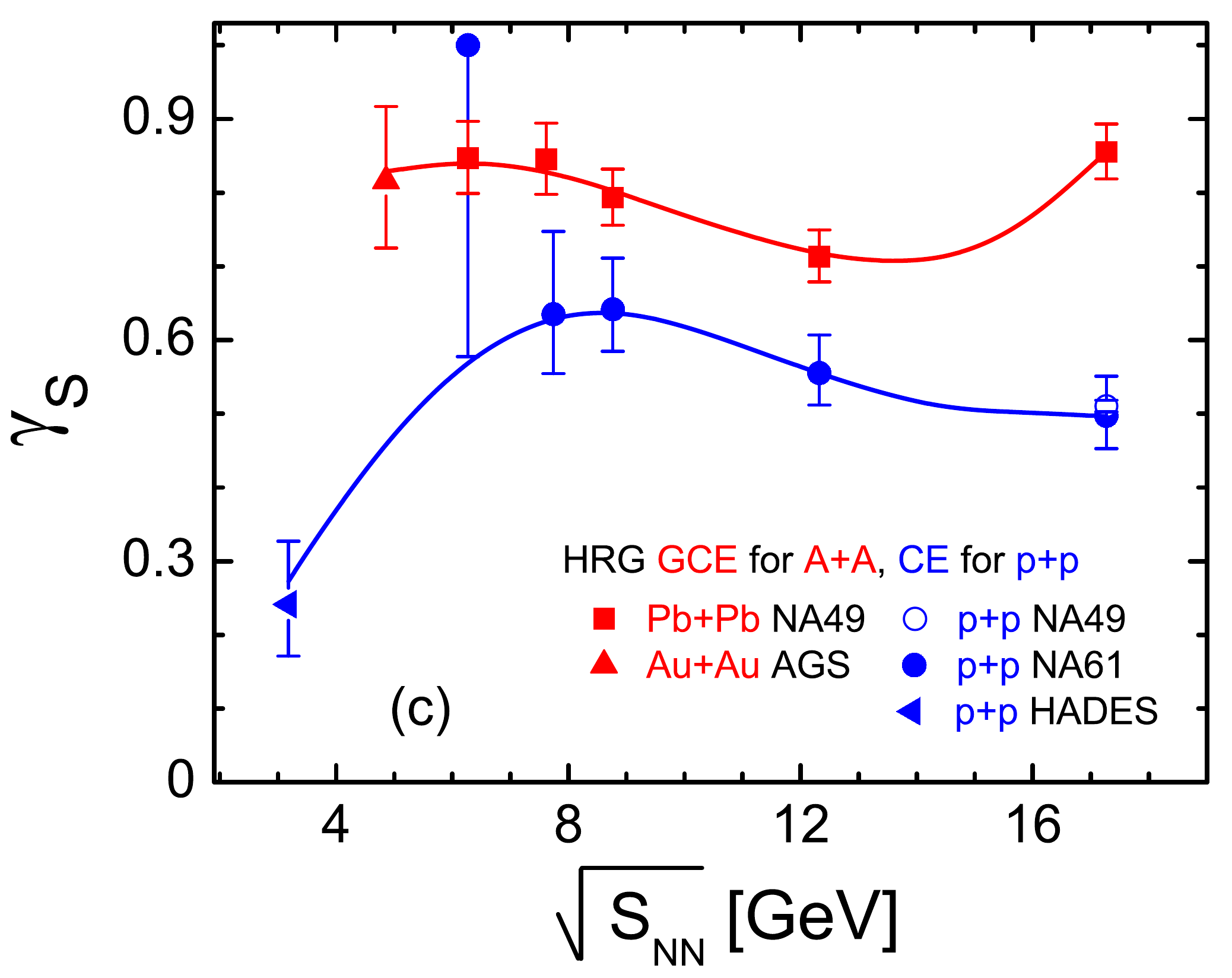}
\includegraphics[width=0.49\textwidth]{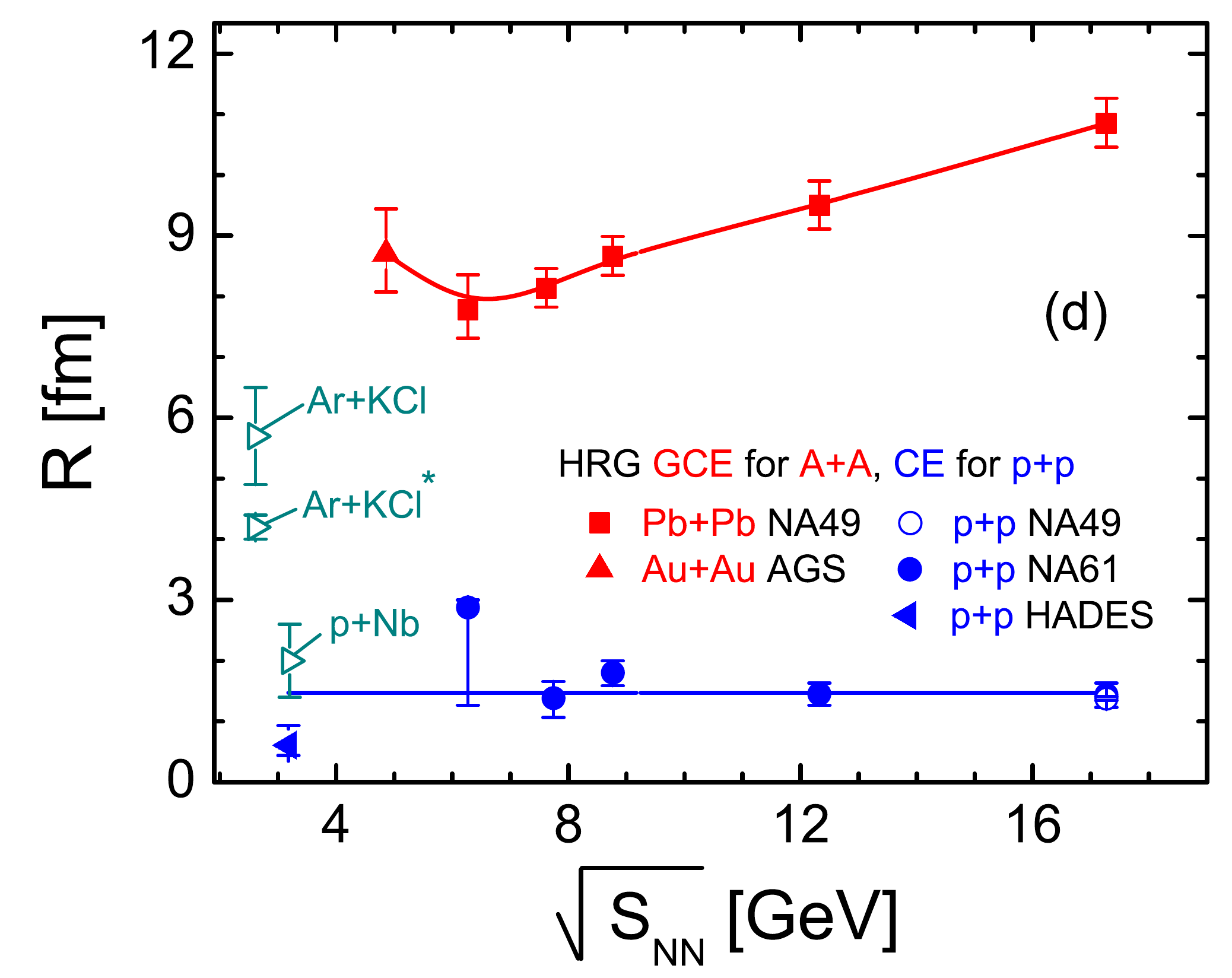}
\caption{ (a) Temperature $T$ as a function of baryon
chemical potential in central Pb+Pb and Au+Au collisions.
Temperature $T$ (b), strangeness saturation factor $\gamma_S$ (c),
radius of the system $R$ (d) in p+p inelastic reactions and
central heavy ion collisions as functions of the collision energy,
see text for more explanations.
%
% The same as (a), zoomed
%in the NA49 energy range.
} \label{fig-T}
\end{figure}

The A+A data at AGS and SPS enegies are fitted within the GCE HRG
model.  The SCE HRG formulation is employed to describe the old
A+A data at SIS (marked as SIS in the figures), and the new data
obtained at SIS by HADES (marked as HADES). The p+p data of HADES
and NA61/SHINE collaborations are analyzed within the CE HRG. The
extracted values of the chemical freeze-out parameters, $T$ and
$\mu_B$, are plotted in Fig.~\ref{fig-T} (a).
The boxes correspond to our fit of the latest compilation of the
NA49 data for central Pb+Pb
collisions~\cite{NA49-1,NA49-3,NA49-2,NA49-latest,NA49-4}.
The full right triangles show our fit to the recent Au+Au data
from HADES~\cite{HADES-A+A}. The open right triangles show the
results of the newest analysis of the p+Nb and Ar+KCl reactions
performed
using THERMUS~3.0 by the HADES collaboration~\cite{Agakishiev:2015bwu}. The
Ar+KCl$^*$ label corresponds to the fit with the reduced number of
fitted yields~\cite{Agakishiev:2015bwu}.
The up and down triangles show our fits to the old Au+Au data
listed in~\cite{Averbeck:2000sn} and in~\cite{Becattini:2000jw}.
%Open circles in
%Fig.~\ref{fig-Cleymans}~(b) correspond to the previous fits from
%the Ref.~\cite{Bec}, that were made for the preliminary data.
%
The Au+Au data at SIS allow to extract temperature and baryonic chemical
potential. They are shown in Figs.~\ref{fig-T} (a) and (b). The
parameters $R$ and $\gamma_S$ however can not be reliably defined, thus,
%there, because either only the particle ratios are measured, or
%the number of measured data is equal to the number of fit
%parameters. Therefore we
they are not shown in Fig.~\ref{fig-T}~(c) and \ref{fig-T}~(d) at
SIS.
%However, the
%corresponding $T(\mu_B)$ points are the most right in
%Fig.~\ref{fig-T}~(a) and define the freeze-out line.
%

The main set of data used in our analysis contains the mean total
multiplicities. All the data from the NA61/SHINE, NA49 at SPS, and
also the point at AGS energy are the total $4\pi$ mean
multiplicities, i.e., the hadron yields integrated over the whole
rapidity range. The p+p data from HADES are also the mean
multiplicities, but extracted from di-electron yields, which may
add some unaccounted systematic error to this point. The SIS Au+Au
data contains the hadron yield ratios and the average number of
participants. The Au+Au data from HADES contains only the ratios
at mid-rapidity.
The HRG model parameters obtained from different sets of data
demonstrate the smooth and consistent behavior. The largest
deviation is seen at the $\sqrt{s_{NN}}=6.3~$GeV in  p+p reactions
from the NA61/SHINE. This can be attributed to the absence of the
antiproton data at $\sqrt{s_{NN}}=6.3~$GeV, which are present in
the measurements at four higher SPS energies provided by the same
collaboration.
%that provides consistent results at four higher
%energies.
%

The grey band in Fig.~\ref{fig-T} (a) is the parametrization from
the Ref.~\cite{Cleymans:2005xv},
 \eq{\label{T-A+A-muB}
 T_{\rm A+A}(\mu_B) ~=~ a ~-~ b\mu_B^2 ~-~ c\mu_B^4~,
 }
where  $a = 0.166 \pm 0.002$~GeV, $b = 0.139 \pm
0.016$~GeV$^{-1}$, and $c = 0.053 \pm 0.021$~GeV$^{-3}$. The width
of the band indicates the corresponding error bars, that were
obtained for each $T(\mu_B)$ point from the errors of the $a$, $b$
and $c$, using the standard methods of propagations of
uncertainties.
Together with the parametrization of the $\mu_B$,
 \eq{\label{muB-Snn}
 \mu_B~=~\frac{d}{1~+~e\,\sqrt{s_{NN}}}~,
 }
where $d=1.308 \pm 0.028$~GeV, $e=0.273 \pm 0.008$~GeV$^{-1}$,
Eq.~(\ref{T-A+A-muB}) allows to plot temperature as the function
of energy.

The fit of our results for $T$ and $\mu_B$ in central
Pb+Pb and Au+Au collisions with the same analytical functions
(\ref{T-A+A-muB}) and (\ref{muB-Snn}) yields somewhat different
parameters, namely: $a=0.157$~GeV, $b=0.087$~GeV$^{-1}$,
$c=0.092$~GeV$^{-3}$, $d=1.477$~GeV, $e=0.343$~GeV$^{-1}$. The
corresponding error bars are rather large and are not shown,
because we used only a few points to determine the freeze-out
line. However, the obtained line in Fig.~\ref{fig-T}~(a) is within
the gray error bars at large chemical potential. Therefore adding
new points there would not change the line. The most important
effect is due to the top SPS points that give smaller temperature
in our analysis.
Interestingly, Eq.~(\ref{T-A+A-muB}) gives $T=157$~MeV at
$\mu_B=0$, which is close to the latest findings at the
LHC~\cite{Stachel:2013zma,Floris:2014pta,Begun:2014aha}.

The change in the parametrization of the chemical freeze-out line
(\ref{T-A+A-muB}) and (\ref{muB-Snn}) is a combination of two
effects: the extension of the list of particles, and the changes
in the experimentally measured particle set. The HRG fit of the
latest NA49 data \cite{NA49-1,NA49-3,NA49-2,NA49-latest,NA49-4} in
the present paper gives approximately constant temperature at top
SPS energies and growing radius of the system, as seen from
Figs.~\ref{fig-T}~(b) and (d). The previous HRG fit \cite{Bec} of
the old NA49 data with a smaller table of particles in the HRG
gave the opposite: constant radius and growing temperature.

In order to further study the effects of heavy resonance decays on
the HRG model parameters, we have analyzed different cuts for the
maximal resonance mass, $M_{\rm cut}$, included in the table of
particles. Varying the cut in the range $1.7<M_{\rm cut}<2.4$~GeV,
we have found that the inclusion of heavy resonances may decrease
the temperature up to 10~MeV, and
%increases the radius of the order of 1~fm.
the effect is stronger for larger collision energy.

The data of the NA61/SHINE on inelastic p+p interactions \cite{NA61-p+p-mult}
are fitted in the CE HRG
model. The experimental results and HRG fit within the CE are
shown in Fig.~\ref{fig-pi} and in Tables~\ref{tab:pp-yields-1} and
\ref{tab:pp-yields-2}.
The obtained $T$, $\gamma_S$, and $R$
parameters are presented in Table~\ref{tab:pp-parameters}. These
parameters are also shown in Figs.~\ref{fig-T} (b-d) for both p+p
and A+A collisions.

In addition, we make the fit of the available p+p data point from
the NA49 at the $E_{\rm
lab}=158$~GeV~($\sqrt{s_{NN}}=17.3~$GeV)~\cite{NA49-p+p,NA49-4}.
The NA49 p+p data include more hadron species, therefore, we check
how the selection of a different particle set influences the
results. We also fit the data from HADES
Collaboration~\cite{HADES-p+p}, for p+p collisions at  $E_{\rm
kin}=3.5$~GeV~($\sqrt{s_{NN}}=3.2~$GeV).

The lower solid line in Fig.~\ref{fig-T} (b) shows the behavior of
temperature in A+A as the function of energy calculated according
to Eqs.~(\ref{T-A+A-muB}) and (\ref{muB-Snn}). Other lines in
Figs.\ref{fig-T} (b)-(d) are the fits that were made to guide the
eye.

The analysis of the NA49 and NA61/SHINE p+p data at $158A$~GeV gives
very close HRG parameters.
%It suggests that the set of hadron
%species measured by the NA61 at this energy is complete.
The corresponding points almost coincide. Larger error bars for
the NA61/SHINE are due to smaller number of measured particles
compared to NA49 (5 versus 18).
The same reason causes much
smaller $\chi^2/N_{\rm dof}$ for NA61/SHINE than that for NA49.
The extracted freeze-out parameters are only slightly changed by
adding the new multiplicity data to the NA61/SHINE set of particles
measured at $158A$~GeV. The further addition of the particles
leads only to an increase of the $\chi^2/N_{\rm dof}$ and decrease
of the error bars, i.e., the larger number of fitted particles
gives more constraints on the range of the HRG parameters.

It is seen from Table \ref{tab:pp-parameters} that values of
$\chi^2 / N_{\rm dof}$ are significantly smaller than unity for
collision energies $\sqrt{s_{NN}}=8.8,~12.3,~17.3~$GeV. This
usually indicates that the experimental errors might be
overestimated.

\begin{table}
\caption{The comparison between fitted and measured total $4\pi$
multiplicities, and the prediction for the unmeasured yields. The
fit is done within the CE formulation of the HRG. Some yields at
the lowest energy are omitted from the table, because the energy
of the system is not enough for their creation. The mean
multiplicities in p+p inelastic interactions are measured by HADES
\cite{HADES-p+p} at $\sqrt{s_{NN}}=3.2$~GeV, and by
NA61/SHINE~\cite{NA61-p+p-mult} at $\sqrt{s_{NN}}=6.3$~GeV and
$7.7$~GeV.
 } %title
 \centering                                                 %centering table
 \begin{tabular*}{\textwidth}{c @{\extracolsep{\fill}} ccccccccc}                                   %columns {format}
 \hline
 \hline
 %Parameters & Values \\
 %\hline
  & \multicolumn{2}{c}{$\sqrt{s_{NN}}=3.2$~GeV } & \hspace{3mm} & \multicolumn{2}{c}{$\sqrt{s_{NN}}$=6.3~GeV } & \hspace{3mm} & \multicolumn{2}{c}{$\sqrt{s_{NN}}=$7.7~GeV } \\
 \cline{2-3}  \cline{5-6} \cline{8-9}                          %horizontal lines
  & Measurement & Fit & & Measurement & Fit & & Measurement & Fit \\
$\pi^+$ &  & $0.782$  &  & $1.582 \pm 0.232$ & $1.698$  &  & $1.985 \pm 0.288$ & $1.903$  \\
$\pi^-$ &  & $0.238$  &  & $1.067 \pm 0.203$ & $0.9345$  &  & $1.438 \pm 0.288$ & $1.174$  \\
$K^+$ &  & $0.00398$  &  & $0.097 \pm 0.015$ & $0.0938$  &  & $0.157 \pm 0.018$ & $0.130$  \\
$K^-$ &  & $3.82 \cdot 10^{-4}$  &  & $0.024 \pm 0.00632$ & $0.0258$  &  & $0.045 \pm 0.005$ & $0.051$  \\
$p$ &  & $1.37$  &  &  & $1.14$  &  &  & $1.1$  \\
$\bar{p}$ &  & --  &  &  & $4.78 \cdot 10^{-5}$  &  & $0.0047 \pm 0.0008$ & $0.0046$  \\
$\Lambda$ &  & $0.00466$  &  &  & $0.0669$  &  &  & $0.0785$  \\
$\bar{\Lambda}$ &  & --  &  &  & $1.57 \cdot 10^{-5}$  &  &  & $0.00161$  \\
$\Sigma^+$ &  & $2.20 \cdot 10^{-4}$  &  &  & $0.0226$  &  &  & $0.0254$  \\
$\bar{\Sigma}^+$ &  & --  &  &  & $3.11 \cdot 10^{-6}$  &  &  & $3.27 \cdot 10^{-4}$  \\
$\Sigma^-$ &  & $6.00 \cdot 10^{-5}$  &  &  & $0.011$  &  &  & $0.0129$  \\
$\bar{\Sigma}^-$ &  & --  &  &  & $4.78 \cdot 10^{-6}$  &  &  & $4.90 \cdot 10^{-4}$  \\
$\Xi^0$ &  & --  &  &  & $9.04 \cdot 10^{-4}$  &  &  & $0.00122$  \\
$\bar{\Xi}^0$ &  & --  &  &  & $7.35 \cdot 10^{-7}$  &  &  & $8.32 \cdot 10^{-5}$  \\
$\Xi^-$ &  & --  &  &  & $6.93 \cdot 10^{-4}$  &  &  & $0.00101$  \\
$\bar{\Xi}^-$ &  & --  &  &  & $8.53 \cdot 10^{-7}$  &  &  & $9.36 \cdot 10^{-5}$  \\
$\Omega$ &  & --  &  &  & $4.06 \cdot 10^{-6}$  &  &  & $1.11 \cdot 10^{-5}$  \\
$\bar{\Omega}$ &  & --  &  &  & $1.70 \cdot 10^{-8}$  &  &  & $3.28 \cdot 10^{-6}$  \\
$\pi^0$ & $0.39 \pm 0.1$ & $0.578$  &  &  & $1.54$  &  &  & $1.76$  \\
$K^0_S$ & $0.0013 \pm 0.0003$ & $0.000977$  &  &  & $0.0501$  &  &  & $0.0811$  \\
$\eta$ & $0.02 \pm 0.007$ & $0.017$  &  &  & $0.0846$  &  &  & $0.134$  \\
$\omega$ & $0.006 \pm 0.002$ & $0.00591$  &  &  & $0.0364$  &  &  & $0.145$  \\
$K^{*+}$ & $(2.0 \pm 0.6) \cdot 10^{-4}$ & $0.000218$  &  &  & $0.0113$  &  &  & $0.0406$  \\
$K^{*-}$ &  & $4.62 \cdot 10^{-5}$  &  &  & $0.00273$  &  &  & $0.0117$  \\
$K^{*0}$ &  & $1.36 \cdot 10^{-4}$  &  &  & $0.00797$  &  &  & $0.0302$  \\
$\bar{K^{*0}}$ &  & $5.86 \cdot 10^{-5}$  &  &  & $0.00347$  &  &  & $0.0144$  \\
%$\rho^+$ &  & $0.0723$  &  &  & $0.0833$  &  &  & $0.241$  \\
%$\rho^-$ &  & $0.0129$  &  &  & $0.0423$  &  &  & $0.122$  \\
%$\rho^0$ &  & $0.0489$  &  &  & $0.071$  &  &  & $0.205$  \\
%$\eta'$ &  & $2.91 \cdot 10^{-4}$  &  &  & $0.00268$  &  &  & $0.0129$  \\
$\phi$ &  & $7.72 \cdot 10^{-5}$  &  &  & $0.00454$  &  &  & $0.0129$  \\
$\Lambda(1520)$ &  & $3.83 \cdot 10^{-5}$  &  &  & $0.00206$  &  &
& $0.00626$
 \end{tabular*}
\label{tab:pp-yields-1}
\end{table}

\begin{table}
\caption{The same as Table~\ref{tab:pp-yields-1} for p+p inelastic
interactions at $\sqrt{s_{NN}}$=8.8, 12.3, and 17.3~GeV measured
by NA61/SHINE~\cite{NA61-p+p-mult}.
 } %title
 \centering                                                 %centering table
 \begin{tabular*}{\textwidth}{c @{\extracolsep{\fill}} ccccccccc}                                   %columns {format}
 \hline
 \hline
 %Parameters & Values \\
 %\hline
  & \multicolumn{2}{c}{$\sqrt{s_{NN}}$=8.8 GeV } & \hspace{3mm} & \multicolumn{2}{c}{$\sqrt{s_{NN}}$=12.3 GeV} & \hspace{3mm} & \multicolumn{2}{c}{$\sqrt{s_{NN}}$=17.3 GeV } \\
 \cline{2-3}  \cline{5-6} \cline{8-9}                          %horizontal lines
  & Measurement & Fit & & Measurement & Fit & & Measurement & Fit \\
$\pi^+$ & $2.221 \pm 0.274$ & $2.383$  &  & $2.556 \pm 0.261$ & $2.618$  &  & $2.991 \pm 0.394$ & $3.161$  \\
$\pi^-$ & $1.703 \pm 0.287$ & $1.603$  &  & $2.030 \pm 0.281$ & $1.844$  &  & $2.494 \pm 0.315$ & $2.368$  \\
$K^+$ & $0.170 \pm 0.025$ & $0.183$  &  & $0.201 \pm 0.014$ & $0.195$  &  & $0.234 \pm 0.022$ & $0.233$  \\
$K^-$ & $0.084 \pm 0.007$ & $0.081$  &  & $0.095 \pm 0.006$ & $0.098$  &  & $0.132 \pm 0.014$ & $0.133$  \\
$p$ &  & $1.05$  &  &  & $1.04$  &  &  & $1.04$  \\
$\bar{p}$ & $0.0059 \pm 0.0007$ & $0.0059$  &  & $0.0183 \pm 0.00186$ & $0.0182$  &  & $0.0402 \pm 0.0033$ & $0.0402$  \\
$\Lambda$ &  & $0.0987$  &  &  & $0.0974$  &  &  & $0.104$  \\
$\bar{\Lambda}$ &  & $0.00196$  &  &  & $0.00553$  &  &  & $0.0108$  \\
$\Sigma^+$ &  & $0.0314$  &  &  & $0.0301$  &  &  & $0.0314$  \\
$\bar{\Sigma}^+$ &  & $4.19 \cdot 10^{-4}$  &  &  & $0.00118$  &  &  & $0.00237$  \\
$\Sigma^-$ &  & $0.018$  &  &  & $0.0179$  &  &  & $0.0202$  \\
$\bar{\Sigma}^-$ &  & $5.90 \cdot 10^{-4}$  &  &  & $0.00164$  &  &  & $0.00315$  \\
$\Xi^0$ &  & $0.00203$  &  &  & $0.00198$  &  &  & $0.0023$  \\
$\bar{\Xi}^0$ &  & $1.02 \cdot 10^{-4}$  &  &  & $2.70 \cdot 10^{-4}$  &  &  & $4.91 \cdot 10^{-4}$  \\
$\Xi^-$ &  & $0.00171$  &  &  & $0.00171$  &  &  & $0.00204$  \\
$\bar{\Xi}^-$ &  & $1.13 \cdot 10^{-4}$  &  &  & $2.95 \cdot 10^{-4}$  &  &  & $5.28 \cdot 10^{-4}$  \\
$\Omega$ &  & $2.31 \cdot 10^{-5}$  &  &  & $2.59 \cdot 10^{-5}$  &  &  & $3.47 \cdot 10^{-5}$  \\
$\bar{\Omega}$ &  & $3.77 \cdot 10^{-6}$  &  &  & $1.03 \cdot 10^{-5}$  &  &  & $1.80 \cdot 10^{-5}$  \\
$\pi^0$ &  & $2.28$  &  &  & $2.53$  &  &  & $3.12$  \\
$K^0_S$ &  & $0.122$  &  &  & $0.137$  &  &  & $0.174$  \\
$\eta$ &  & $0.173$  &  &  & $0.204$  &  &  & $0.256$  \\
$\omega$ &  & $0.185$  &  &  & $0.253$  &  &  & $0.343$  \\
$K^{*+}$ &  & $0.0519$  &  &  & $0.0647$  &  &  & $0.0804$  \\
$K^{*-}$ &  & $0.019$  &  &  & $0.026$  &  &  & $0.0377$  \\
$K^{*0}$ &  & $0.0403$  &  &  & $0.0517$  &  &  & $0.0665$  \\
$\bar{K^{*0}}$ &  & $0.0227$  &  &  & $0.0305$  &  &  & $0.0432$  \\
%$\rho^+$ &  & $0.283$  &  &  & $0.362$  &  &  & $0.458$  \\
%$\rho^-$ &  & $0.166$  &  &  & $0.216$  &  &  & $0.295$  \\
%$\rho^0$ &  & $0.25$  &  &  & $0.321$  &  &  & $0.415$  \\
%$\eta'$ &  & $0.0156$  &  &  & $0.0219$  &  &  & $0.0292$  \\
$\phi$ &  & $0.0157$  &  &  & $0.0183$  &  &  & $0.0211$  \\
$\Lambda(1520)$ &  & $0.00707$  &  &  & $0.00853$  &  &  &
$0.00978$
 \end{tabular*}
\label{tab:pp-yields-2}
\end{table}

\begin{table}
 \caption{Summary of the fitted parameters in p+p inelastic interactions
  within CE formulation of HRG.} %title
 \centering                                                 %centering table
 \begin{tabular*}{\textwidth}{c @{\extracolsep{\fill}} cccc}                                   %columns {format}
 \hline
 \hline
 %Parameters & Values \\
 %\hline
 Parameters & $\sqrt{s_{NN}}$=3.2 GeV & $\sqrt{s_{NN}}$=6.3 GeV & $\sqrt{s_{NN}}$=7.7 GeV\\
 \hline                                             %horizontal lines
 $T$ (MeV) & $141.0^{+15.9}_{-19.3}$ & $102.0^{+51.6}_{-2.8}$ & $154.6^{+13.6}_{-8.9}$ \\
 $\gamma_S$ & $0.242^{+0.086}_{-0.071}$ & $1.000^{+0.000}_{-0.423}$ & $0.635^{+0.112}_{-0.081}$\\
 $R$ (fm) & $0.61^{+0.32}_{-0.17}$ & $2.88^{+0.12}_{-1.61}$ & $1.38^{+0.28}_{-0.32}$ \\
 $\chi^2/N_{\rm dof}$ & $4.95/2$ & $0.80/1$ & $4.44/2$
 %\\
 %$\chi^2_{\rm stat}/N_{\rm dof}$ &   & $4.22/1$ & $9.53/2$ \\
 \\
 \hline
  & $\sqrt{s_{NN}}$=8.8 GeV & $\sqrt{s_{NN}}$=12.3 GeV & $\sqrt{s_{NN}}$=17.3 GeV\\
 \hline
 $T$ (MeV) & $144.1^{+5.9}_{-4.8}$ & $163.3^{+7.9}_{-6.2}$ & $170.8^{+8.0}_{-7.0}$  \\
 $\gamma_S$ & $0.642^{+0.069}_{-0.058}$ & $0.555^{+0.052}_{-0.043}$ & $0.497^{+0.053}_{-0.045}$\\
 $R$ (fm) & $1.80^{+0.20}_{-0.21}$ & $1.46^{+0.18}_{-0.19}$ & $1.44^{+0.19}_{-0.21}$ \\
 $\chi^2/N_{\rm dof}$ & $0.89/2$ & $0.88/2$ & $0.35/2$
 %\\
 %$\chi^2_{\rm stat}/N_{\rm dof}$ & $2.00/2$ & $1.47/2$ & $0.56/2$
 \end{tabular*}
\label{tab:pp-parameters}
\end{table}

The temperature in p+p is gradually increasing with collision
energy from $T_{\rm p+p}\simeq 130$~MeV to $T_{\rm p+p}\simeq
170$~MeV. The sudden drop of the temperature at
$\sqrt{s_{NN}}=6.3~$GeV is correlated with the corresponding
increase of the radius $R_{\rm p+p}$ and the $\gamma_S$. Large
error bars at this energy indicate that the measurement of the
$\bar{p}$ and/or other (anti)baryon is needed to constrain the
parameters.

The HRG model for multiplicities in p+p reactions at
$\sqrt{s_{NN}}=19.4$~GeV was considered in
Ref.~\cite{CE2,Becattini:1997rv}. This energy is close to the top
SPS energy. The p+p temperature in
Ref.~\cite{CE2,Becattini:1997rv} is in agreement with our results
within the error bars. The e$^+$+e$^-$ and p+$\bar{\rm p}$
temperatures in Ref.~\cite{CE2,Becattini:1997rv} are also close to
our results. The p+p temperature for $\sqrt{s_{NN}}=200$~GeV at
RHIC was found to be $T_{\rm p+p}\simeq
170$~MeV~\cite{Becattini:2010sk}, which is in agreement with a
slow increase and a saturation of the temperature obtained in our
fit.

A possible universal mechanism of thermal hadron production in
collisions of elementary particles was suggested in
Ref.~\cite{Castorina:2007eb}. It connects the temperature to the
string tension between quarks, and explains why the temperatures
in e$^+$+e$^-$, p+p, and p+$\bar{\rm p}$ appear to be close to
each other. On the other hand, secondary collisions and medium
effects are evidently important in central Pb+Pb (or Au+Au)
collisions.

The unexpected finding is a decrease of $\gamma_S$ parameter with
collision energy in p+p inelastic reactions in the SPS energy
region. Together with the point from HADES one may conclude that
$\gamma_S$ increases at small energies and probably has a maximum
at the low SPS energy. As seen from Fig.~\ref{fig-T} (c) a similar
behavior is observed for $\gamma_S$ in central Pb+Pb and Au+Au
collisions. Our results at the top SPS energy agree with those
obtained in Ref.~\cite{CE2,Becattini:1997rv}. We note that
previous studies on hadron production in p+p collisions dealt with
higher collision energies than those considered in our paper.
There the $\gamma_S$ parameter was generally found to increase
with collision energy. Our results at SPS energies are also in a
slight contrast with recent paper~\cite{Castorina:2016eyx}, where
it is implied that $\gamma_S$ universally increases with collision
energy. However, the corresponding error bars are still large to
make the final conclusions.

As seen from Fig.~\ref{fig-T} (d) the system radius in p+p
inelastic reactions is approximately independent of the collisions
energy, $R_{\rm p+p}\simeq 1.5$~fm. An exception is the lowest
energy p+p point from HADES. The volume that was found in the p+p
reactions at RHIC gives essentially larger values of the radius,
$R_{\rm p+p}\simeq 3.6$~fm~\cite{Becattini:2010sk}. We do not see,
however, the increase of $R$ at the SPS energies.
%It means that the fireball radius starts to
%grow in p+p collisions at energies that are much larger than the
%SPS.
%
The dependence of the radius on the collision energy is
rather different in p+p and A+A collisions at the SPS
energies: it grows in central A+A collisions,
while in p+p inelastic reactions  the radius is approximately
constant.

Note that the excluded volume corrections~\cite{EV1} neglected in
the present paper do not change the results for the intensive HRG
model parameters -- $T, ~\mu_B, ~\gamma_S$ -- only if all
hard-core radii of hadrons are assumed to be equal to each other.
However, the excluded volume corrections can significantly reduce
the densities~\cite{Begun:2012rf} and, thus, increase the total
system volume at chemical freeze-out in a comparison to the ideal
HRG. Therefore, the  finite size of hadrons influences the total
system volume: the values of $R_{\rm p+p}$ and $R_{\rm A+A}$ would
become larger and their energy dependence would be changed.
The intensive HRG model parameters can also be influenced, if one
considers hadrons with different hard-core
radii~\cite{EV2,Vovchenko:2015cbk}.
However, this will require
additional assumptions (and new model parameters) about sizes of various hadrons,
which are presently rather poorly constrained.

%Additionally, for $p+p$ collisions
%such analysis requires the
%excluded-volume model formulation for HRG in canonical ensemble,
%which is presently lacking. For these reasons the excluded-volume effects
%are not considered in the present study.
%Unfortunately, there is no
%enough data to provide such an analysis. However, the continuous
%effort to measure more and more particle multiplicities by
%different collaborations gives a hope that this analysis will be
%done.
%

Thermal HRG model parameters for all intermediate systems like p+A
or A+A collisions of small nuclei are expected to be in between
those found in the present study for Pb+Pb and p+p reactions.
Presently existing results, particularly, the independent analysis
of p+Nb and Ar+KCl reactions by HADES
Collaboration~\cite{Agakishiev:2015bwu} supports this statement.
The existing results also indicate that temperatures reached by
different systems in the beam energy scan at the SPS might be very
similar. However, exact conservation of net strangeness and
baryonic number remains important in p+A and light nuclei
collisions. Therefore, the total number of strange hadrons and
antibaryons per nucleon participant may be essentially reduced for
small systems despite of their larger temperatures.

\section{Summary}\label{sum}

Our analysis of the new data for mean hadron multiplicities
demonstrates that both transport models -- UrQMD and HSD -- should
be significantly modified and tuned to the presently available p+p
data at SPS energies. This is indeed important as the properties of p+p reactions
are used in these transport models as the input for Monte Carlo
simulations of A+A collisions.

The CE HRG model leads to the good description of the
data on hadron multiplicities in p+p interactions.
Our results define the range of the chemical freeze-out parameters
-- $T$, $\gamma_S$, and $R$ -- that can be reached in collisions
of different size nuclei during the energy and system size
scanning at the SPS energy range.
The comparison of the obtained HRG parameters in p+p inelastic
reactions and central Pb+Pb (or Au+Au) collisions shows that the
freeze-out temperature in p+p is larger than that in A+A, $T_{\rm
p+p}>T_{\rm A+A}$. The temperature in p+p slowly grows with energy
from $130$ to $170$~MeV, while the A+A temperature strongly
increases at small collision energy and saturates fast at $T_{\rm
A+A}\simeq157$~MeV, in contrast to $T_{\rm A+A}\simeq166$~MeV
found in previous studies.
% because $T_{\rm A+A}$ is additionally governed
%by baryon chemical potential, $\mu_B$, and follows the A+A
%freeze-out line.
%
In the considered energy range the largest difference $T_{\rm
p+p}-T_{\rm A+A}\cong 60~$MeV is at low energies. The $T_{\rm
p+p}\simeq T_{\rm A+A}$ at $\sqrt{s_{NN}}=6.3-7.7~$GeV, and then
the difference grows again reaching about 20~MeV at the highest
SPS energy.

At all collision energies the $\gamma_S$ parameter in central
Pb+Pb and Au+Au collisions is larger than that in p+p inelastic
reactions. It seems that in both cases this parameter has a
%maximal value
local maximum at the low SPS energy. The obtained results also
indicate that $\gamma_S$ parameter in p+p interactions decreases
with collision energy at SPS energies. While the error bars are
still too large to make firm conclusions, this is in contrast with
the previous studies, which dealt with higher collision energies,
and predicted a monotonous increase of the $\gamma_S$ with
the collision energy.

The dependence of the system radius on the collision energy
is rather different in  central Pb+Pb collisions and p+p reactions
in the SPS energy region. The radius $R_{\rm A+A}$ increases
with collision energy for 40\%, while $R_{\rm p+p}$ has
approximately constant value. The $R_{\rm A+A}$ dependance
found in our analysis is different than in the previous
studies\footnote{It should be noted that finite size of hadrons,
neglected in the present paper, may have an influence on the total
system volume and its dependence on the collision energy.}, where
$R_{\rm A+A}$ was approximately constant at the SPS. The radius,
temperature, and the $\gamma_S$ parameters in p+p reactions at such
low collision energies are obtained for the first time.
%

%
%The measurements by HADES determine the behavior of the parameters
%for p+p and A+A collisions in the region, where they differ the
%most. Therefore, a better estimation of the error bars for the
%mean multiplicities in p+p, and the measurement of the total
%multiplicities, instead of the current midrapidity ratios, in
%Au+Au by HADES would be of the great importance.
%

The fit of the mean multiplicities considered in the present
paper, both in p+p and A+A reactions, assumes that a system
behaves at the chemical freeze-out as the ideal hadron resonance
gas. Thus, the effects of the possible deconfinement phase
transition %and/or QCD critical point
may be signaled as some irregular behavior of the obtained
parameters and deviations of the data from the HRG model results.
 We do see an indication of such an irregular behavior for
$\gamma_S$ at low energies in A+A collisions and, surprisingly,
even stronger in p+p interactions.
However, there is no enough data at low energy A+A, while the
lowest available p+p point contains a different set of measured
particles than for other p+p points.
%
%The existing p+p data is described well by the CE HRG model,
%however the extracted thermal parameters at freeze-out
%the chemical freeze-out
%temperature
%{\bf have rather large uncertainties, much larger than for A+A.
Therefore, the uncertainties in extracted parameters are still too
large to make firm conclusions and
%Thus,
more data in both A+A and p+p are needed at this energy range
in order to clarify this point.

The measurements of total particle multiplicities for a
wider set of hadron species are needed.
The minimal set of fitted multiplicities should include particles
possessing all three conserved charges -- $B, ~S, ~Q$ -- and the
corresponding anti-particles for both p+p and A+A.
%antibaryons, and the number of participants.
For example, an appropriate set of hadron species may include
$\pi^+$, $\pi^-$, $K^+$, $K^-$, $p$, and $\bar{p}$. Therefore, the
additional measurements of anti-proton at the lowest SPS and
proton mean multiplicities in both p+p and intermediate A+A
reactions at all SPS energies are necessary.

%{\bf The
%measurements} of the event-by-event fluctuations of hadron yields
%would also be useful {\bf to answer the question about the
%deconfinement phase transition, because fluctuations} are
%determined by the finer details of the underlying equation of
%state as compared to the hadron multiplicities.

%
%The most promising way to search for these signals is the
%simultaneous analysis of the data on the mean hadron
%multiplicities and their event-by-event fluctuations. The
%systematic data on hadron yields and fluctuations in
%nucleus-nucleus collisions at the SPS energies will be soon
%available.

%
%It is especially important at energies below
%$\sqrt{s_{NN}}\sim10~$GeV, where one expects the signal of the
%onset of deconfinement.}

\begin{acknowledgments}
We are thankful to M. Gazdzicki and S. Pulawski for fruitful
comments, and to J. Cleymans and S. Wheaton for the clarification
of decay contributions to the hadron yields in THERMUS. We thank
NA49, NA61/SHINE, and HADES Collaborations for providing us the
data.
V.V.B. thanks for support by Polish National Science Center grant
No. DEC-2012/06/A/ST2/00390. The work of M.I.G. was supported by
the Program of Fundamental Research
%of the Department of Physics
%and Astronomy
of the National Academy of Sciences of Ukraine.
\end{acknowledgments}

\end{document}